\documentclass[%
 aip,
 apl,%
 amsmath,amssymb,
reprint,%
]{revtex4-1}

\usepackage{color, graphicx}% Include figure files
\usepackage{dcolumn}% Align table columns on decimal point
\usepackage{bm}% bold math
\usepackage{amssymb}
\usepackage{amsmath}
\usepackage{latexsym}
\usepackage{amsfonts}
\usepackage{amsmath}
\usepackage{multirow}
\usepackage{ifthen}
\DeclareMathOperator{\sech}{sech}

\begin{document}

\preprint{}

\title{Wave focusing using symmetry matching in axisymmetric acoustic gradient index lenses}

\author{V. Romero-Garc\'ia}
 \altaffiliation{Instituto de Investigaci\'on para la Gesti\'on Integrada de zonas Costeras, Universitat Polit\`ecnica de Val\`encia, Paranimf 1, 46730, Grao de Gandia, Val\`encia, Spain}
 \affiliation{LUNAM Universit\'e, Universit\'e du Maine, CNRS, LAUM UMR 6613, Av. O. Messiaen, 72085 Le Mans, France}

\author{A. Cebrecos}
\author{R. Pic\'o}
\author{V.J. S\'anchez-Morcillo}
 \affiliation{Instituto de Investigaci\'on para la Gesti\'on Integrada de zonas Costeras, Universitat Polit\`ecnica de Val\`encia, Paranimf 1, 46730, Grao de Gandia, Val\`encia, Spain}
 
\author{L.M. Garcia-Raffi}
\affiliation{Instituto Universitario de Matem\'atica Pura y Aplicada. Universitat Polit\`ecnica de Val\`encia, Camino de Vera s/n, 46022, Val\`encia, Spain.}

\author{J.V. S\'anchez-P\'erez}
\affiliation{Centro de Tecnolog\'ias F\'isicas: Ac\'ustica, Materiales y Astrof\'isica. Universitat Polit\`ecnica de Val\`encia, Camino de Vera s/n, 46022, Val\`encia, Spain.}

%\date{\today}

\begin{abstract}

The symmetry matching between the source and the lens results of fundamental interest for lensing applications. In this work we have modeled an axisymmetric gradient index (GRIN) lens made of rigid toroidal scatterers embedded in air considering this symmetry matching with radially symmetric sources. The sound amplification obtained in the focal spot of the reported lens (8.24 dB experimentally) shows the efficiency of the axisymmetric lenses with respect to the previous Cartesian acoustic GRIN lenses. The axisymmetric design opens new possibilities in lensing applications in different branches of science and technology.

\end{abstract}

\pacs{43.20.Fn, 43.20.Gp, 43.20.Mv, 63.20.-e}
%\keywords{Phononic crystal, Sonic Crystal, Complex band structures, Evanescent modes, Waveguides}

\maketitle

Photonic\cite{John, Yablonovitch} and phononic\cite{Kushwaha93, Martinez95} crystals have been revealed in the last years as promising alternatives to control the propagation of electromagnetic and acoustic waves respectively and, based on new physical concepts, with extensive applications in both optics\cite{joannopoulos08} and acoustics\cite{Pennec10}. Depending on the ratio between the wavelength of the incident wave, $\lambda$, and the lattice constant of the crystals, $a$, the basic mechanism describing the action of the crystal on the wave can be best interpreted in terms of refraction\cite{Cervera02} or diffraction\cite{Sanchez98}. In the long wavelength regime, i.e., $\lambda>>a$, crystals can be considered as homogeneous materials with effective properties\cite{Sheng95, Mei06}, therefore one can design refractive\cite{Cervera02} or gradient index (GRIN) \cite{Lin2009} lenses to control waves. In this direction, metamaterial acoustic GRIN lenses have recently been designed by using unit cells based on cross-shape scatterers \cite{Zigoneanu2011} and on coiling up space \cite{Li2012}, providing a high transmission efficiency and small size. On the other hand, the case $\lambda\simeq a$ corresponds to diffractive regime, where the crystal is strongly dispersive. Yang \textit{et al.}\cite{Yang04} reported the first three dimension (3D) phononic crystal showing the focusing of ultrasonic waves in this regime. Since then several phononic lenses have been designed by using the curvature properties of the isofrequency contours, making use of the all angle negative refraction\cite{Luo02} and the convex isofrequency contours\cite{Ke05}.

\begin{figure}
%\begin{center}
\includegraphics[width=85mm]{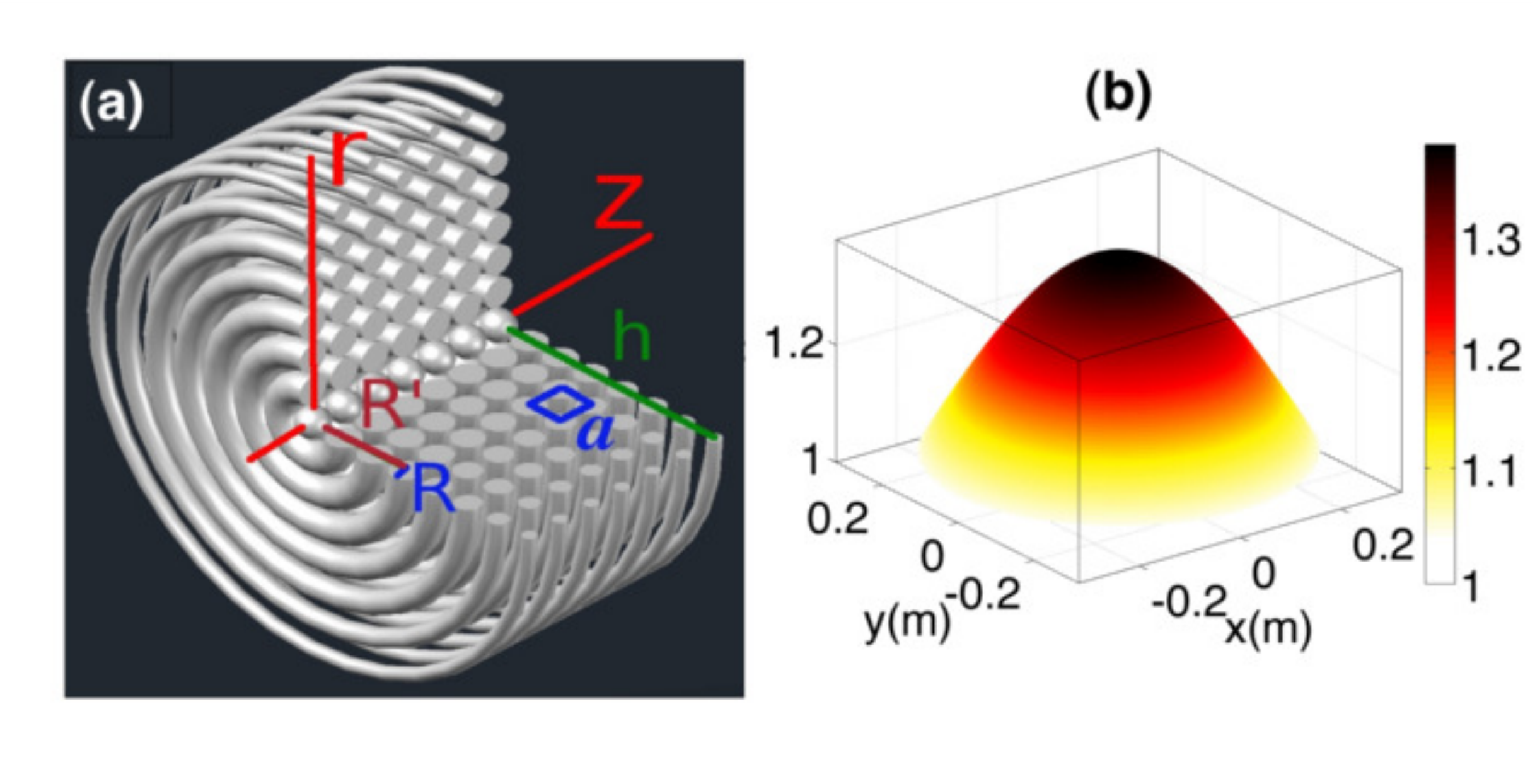}
%\end{center} %\begin{quote}
\caption{(Color online) (a) Parameters used in the axisymmetric GRIN lens. (b) Axysimmetric distribution of the refraction index used in the axisymmetric GRIN lens.}
 \label{fig:fig1}
%\end{quote} 
\end{figure}

In most of the practical situations the sound wave sources have radial symmetry. Examples can be found in domains as aeroacoustics, microfluidics or medical ultrasound. In this situation the symmetry of the lens becomes relevant and one should consider the full source-lens system in order to improve the efficiency of the joint focusing device. Most of the focusing mechanisms described above have been conceived for cartesian lenses (those presenting translational symmetry, as for example a squared array of cylinders), which do not match with the radial symmetry of the source. A cartesian lens in general match with a semi-infinite rectangular radiating surface, which in the asymptotic limits, corresponds to a plane (radiating an unbounded plane wave) or to a line (radiating a cylindrical beam). %The axisymmetric lenses however present a symmetry matching with radial symmetric sources as, for example, a circular radiating piston, whose asymptotic limits are the infinite radiating plane and the point source. Therefore, in order to have an optimal control of the radiated beams, a symmetry matching between the lens and the source is needed.
The axisymmetric lenses however present a symmetry matching with radial symmetric sources as, for example, the circular radiating piston. The asymptotic limits of this circular radiating piston are the infinite radiating plane (radiating an unbounded plane wave) and the point sources. 

Some recent works introduce axisymmetric discrete systems\cite{Chang2012, Sanchis2010} with the aim of focalizing optical or acoustical waves. In the refractive regime, a transformational design of an axial symmetric three-dimensional GRIN lens was theoretically proposed in Ref. [\onlinecite{Chang2012}]. On the other hand, in the diffractive regime, in Ref. \cite{Sanchis2010} the authors propose to maximize the focusing properties in a 2D system of cylindrical rigid scatterers embedded in air and obtain, by rotation of the optimized structure, an axial symmetric lens formed by rigid toroidal scatterers. The structure was validated experimentally and the obtained sound amplification in the focus was remarkably high, showing that the rotational symmetry of the system increases its efficiency. However, the structure was designed and optimized for a cartesian system, where the wave equation is different than that of the axisymmetric case. In the axisymmetric situation the equation presents a term proportional to $1/r$, making the symmetry axis of singular relevance.

In this work we propose the model and the experimental realization of an axisymmetric GRIN lens working with a circular piston source radiating Gaussian beams in the long wavelength regime ($\lambda>4a$). In this range of frequencies the axisymmetric lens can be considered as an equivalent fluid. Due to the symmetry matching between the radiated beam and the GRIN lens, a high sound level is found in the focus spot. We  have characterized the focusing properties of the complete system, demonstrating values of the sound amplification (gain) higher than those obtained previously with acoustic GRIN lenses. The lens is modeled in the axial plane (horizontal plane) without loss of generality as shown in Fig. \ref{fig:fig1}(a), and it is made of rigid toroidal scatterers embedded in air. Each scatterer is represented by a major radius, $R'$, and a minor radius, $R$. In the axial plane the distance between neighbor scatterers is $a$, forming a square array as shown in Fig. \ref{fig:fig1}(a). 

\begin{figure}
%\begin{center}
\includegraphics[width=8.5cm]{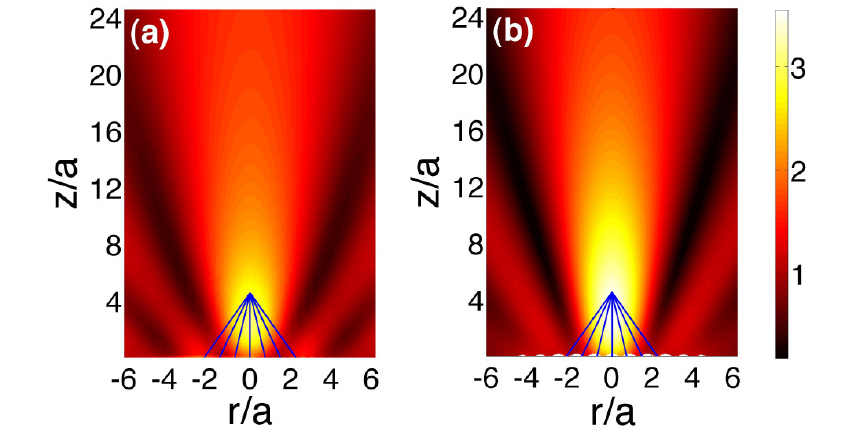}
%\end{center} %\begin{quote}
\caption{(Color online) Comparison between the acoustic field (absolute value of pressure, $|p|$) behind the equivalent fluid lens (a) and the axisymmetric GRIN lens (b) at $\lambda=4.25a$. Blue lines represent the ray-tracing trajectories calculated from the derivative of Eq. (\ref{eq:foc}) at the interface of the GRIN lens and then considering Snell's law of refraction.}
 \label{fig:fig2}
%\end{quote} 
\end{figure}

In the long wavelength regime, the minor radius of each scatterer can be selected to fix the filling fraction, $f(r)=\pi R(r)^2/a^2$, at a position $r$ from the center of the lens\cite{Cervera02}, and the index of refraction, $n(r)$, can be written in terms of $f$ as\cite{Cervera02, Krokhin03}
\begin{equation}
n(r)=\frac{c_{host}}{c_{eff}} =\sqrt{1+ f(r)}. \label{index}
\end{equation}
Then, by a gradual change of the filling fraction we can design a refraction index profile in the vertical plane of the lens, perpendicular to the axial $z$-direction. In this work we use the hyperbolic secant profile that has been proved to reduce the aberration of the focal spot\cite{Gomez02}, defined as
\begin{equation}
n(r)= n_{0}\sech(\alpha r)  \label{profile},
 \end{equation}
where $n_{0}=n(r=0)$ is the refractive index on the $z$-axis ($r=0$) and $\alpha$ is the gradient coefficient, 
\begin{equation}
     \alpha=\frac{1}{h}\cosh^{-1}\left(\frac{n_0}{n_h}\right) \label{alpha},
 \end{equation}
with $h$ the half-height of the lens, and $n_{h}$ the refraction index at the lens edges $(r=\pm h)$. Figure \ref{fig:fig1}(b) shows the hyperbolic secant refractive index profile, where we have selected $n_0=1$ and $n_h=1.33$, being $h=7a$ the thickness of the lens, so we have used a sample with 7 planes of toroidal scatterers as shown in Fig. \ref{fig:fig1}(a).

\begin{figure}
%\begin{center}
\includegraphics[width=8.5cm]{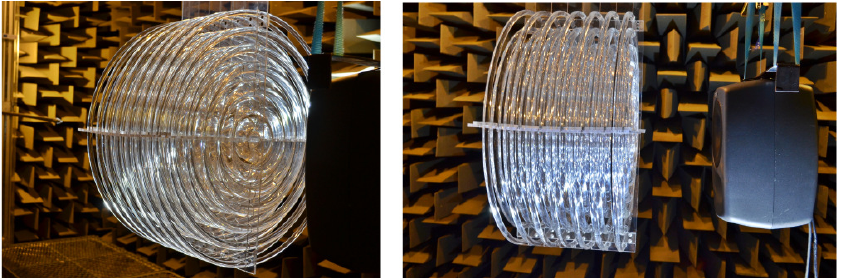}
%\end{center} %\begin{quote}
\caption{(Color online) Views of the experimental set-up showing the lens-source system.}
 \label{fig:fig3}
%\end{quote} 
\end{figure}

\begin{figure*}
%\begin{center}
\includegraphics[width=150mm]{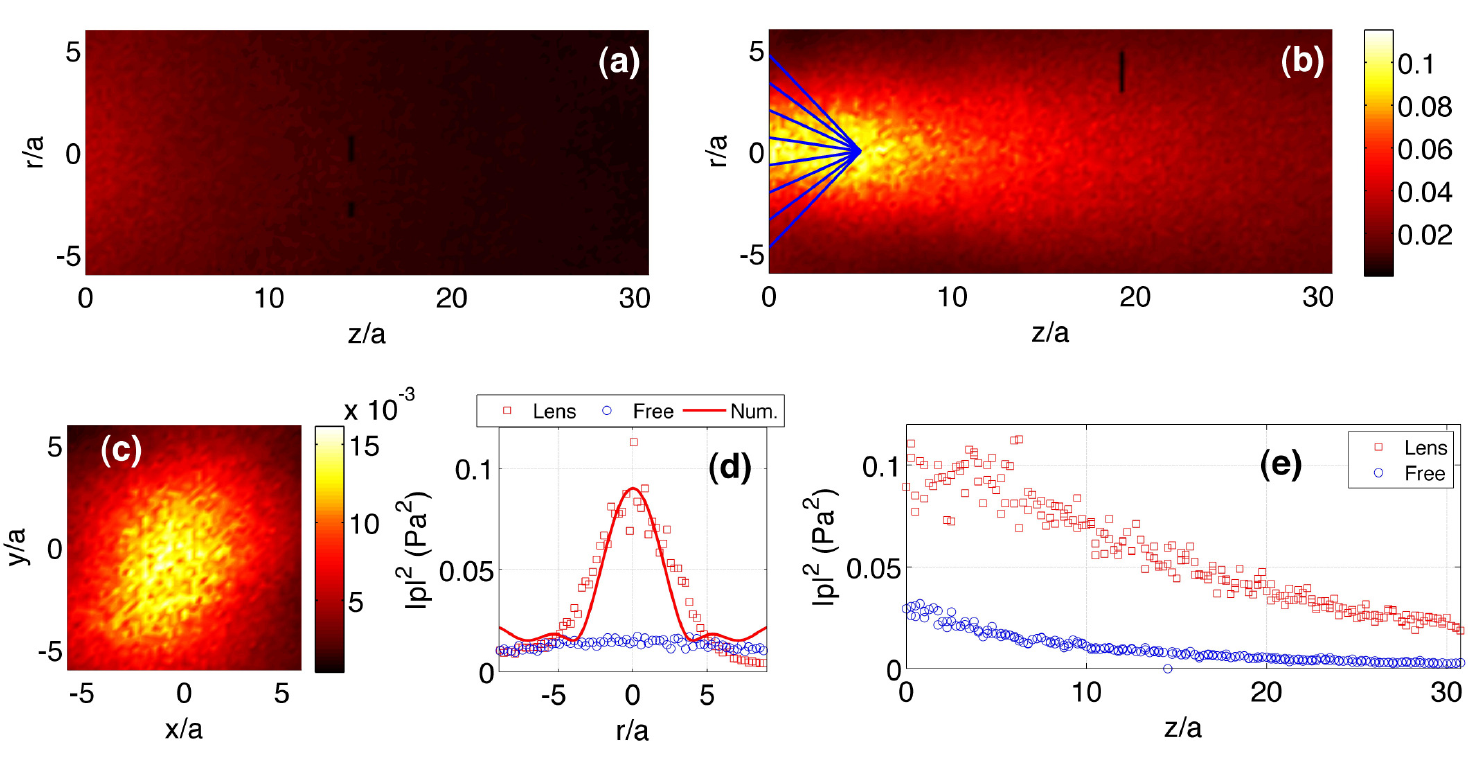}
%\end{center} %\begin{quote}
\caption{(Color online) Experimental results of the axisymmetric GRIN lens. (a) and (b) show the experimental intensity axial map ($|p|^2$) without and with the lens respectively at $\lambda=4.4a$. Blue lines in (b) show the focal spot position obtained by using the ray-tracing calculations. (c) shows the experimental intensity radial map at the position of the focus. (d) and (e) show the transversal (at $z/a=5$) and longitudinal (at $r/a=0$) cuts respectively. Blue open circles (red open squares) show the intensity in free field (behind the axisymmetric GRIN lens). As an eye-guide for comparison the red continuous line represents the numerical simulation results obtained using FEM.}
 \label{fig:fig4}
%\end{quote} 
\end{figure*}

The numerical results are obtained by solving the acoustic wave equation using the finite element method (FEM) applied in an axisymmetric domain surrounded by perfectly matched layers, in order to simulate the Sommerfeld radiation conditions. The scatterers are assumed acoustically rigid (infinite impedance), corresponding to Neumann boundary conditions at the interface with the fluid. The incident field was generated by a circular piston of diameter D = 5a placed at a distance of 5a from the source. Figure \ref{fig:fig2} shows the comparison of the acoustic field behind an axisymmetric slab of thickness $7a$ made of the equivalent fluid medium with the index profile defined by Eq. (\ref{profile}) (see Fig. \ref{fig:fig2}(a)) and the acoustic field behind the real axisymmetric structure made with toroids shown in Fig. \ref{fig:fig1}(a) at $\lambda=4.25a$ (see Fig. \ref{fig:fig2}(b)). In both cases, the hyperbolic secant refractive index profile was used to determine the ray-tracing trajectory within GRIN lens as follows\cite{Lin2009}:
\begin{equation}
y(x)=\frac{1}{h}\sinh^{-1}(u_0H_f(x)+\frac{\mathrm d{u}_0}{\mathrm dx}H_a(x)), \label{eq:foc}
\end{equation}
where $u_0=\sinh y_0$; $H_a(x)=\sin(\alpha x)/\alpha$ and $H_f(x)=\cos(\alpha x)$. Taking the derivative of Eq. (\ref{eq:foc}) at the interface of the GRIN lens and considering Snell's law of refraction one can obtain the focus position behind the GRIN lens. Blue continuous lines in Fig. \ref{fig:fig2} represent the ray-tracing trajectories behind the samples for the case considered in this work, showing the focal point at $z\simeq5a$. A good agreement between simulations for the equivalent fluid, Fig. \ref{fig:fig2}(a), and the real structure, Fig. \ref{fig:fig2}(b), is observed.

An experimental set-up was designed to characterize the focusing properties of the axisymmetric GRIN lens in an anechoic chamber and to obtain quantitative data of the acoustic field behind the lens. The dimensions of the echo-free chamber are $8\times 6\times 3$ m$^3$. The automatized acquisition system 3DReAMS (3D Robotized e-Acoustic Measurement System)\cite{Romero10a} was used to scan the acoustic field distribution. Both the source and the lens were hanged and accurately oriented. We notice that the system is based on the axisymmetric properties, so the alignment between the source and the lens is found critical to experimentally obtain a high sound level in the focal spot. Figure $\ref{fig:fig3}$ shows two views of the experimental set-up. The toroidal scatterers of the lens are made of plexiglass, which acoustic impedance is $\sim 6000$ times bigger than that of the air. Therefore, the toroidal scatterers can be considered acoustically rigid. A loudspeaker is excited with a white noise signal. The diameter of the circular source is $D=5a$ and it is placed at a distance of $5a$ from the lens. In this work we present all the results in normalized units with respect to the lattice constant of the sample, so it is worth noting that for our experiments we have used $a=4$ cm.

The acoustic axisymmetric GRIN lens has been designed to operate as a focusing device in the homogenization regime. The lens presents a broadband behavior in the low frequency range, with modulations in amplitude due to the Fabry-P\'erot resonances of the slab. Here we evaluate the intensity maps ($|p|^2$) at $\lambda=4.4a$. We measured the case of free propagation (Fig. \ref{fig:fig4}(a)) and the case of the propagation through the GRIN lens in the axial (Fig. \ref{fig:fig4}(b)) and the radial planes (\ref{fig:fig4}(c))). As in Fig. \ref{fig:fig2}, the blue continuous lines in Fig. \ref{fig:fig4}(b)  represent the ray-tracing trajectory behind the sample showing the focal spot at $z\simeq5a$ in good agreement with the predictions shown in Fig. \ref{fig:fig2}.

Figures \ref{fig:fig4}(d) and \ref{fig:fig4}(e) represent the experimental transversal (at $z/a=5$) and longitudinal (at $r/a=0$) cross-sections respectively. Blue open circles and red open squares show the intensity in free propagation and behind the axisymmetric GRIN lens respectively. One can observe a considerably gain in the symmetry axis with respect to the case of free field propagation. Red continuous line in Fig. \ref{fig:fig4}(d) represents the numerical transversal cut obtained using FEM. Using the data of Fig. \ref{fig:fig4}(d) we can quantitatively characterize the sound amplification (SA) produced in the focal point as well as the Full Width at Half Maximum (FWHM). From Fig. \ref{fig:fig4}(d) the FWHM$=1.04\lambda$. Considering as a reference pressure the value of the experimental measurement in free field, $|p|^2_{free}=0.013$ Pa, we can evaluate the sound amplification in the focal point as $SA(dB)=10\log_{10}(|p|^2_{lens}/|p|^2_{free})=8.24$ dB. The high value of $SA$ for the case of the axisymmetric lens obtained in this work is in contrast with the value of the $SA$ obtained using cartesian lenses\cite{Climente2010, Martin2010}. In the case of the cartesian lenses the focus is extended over the third dimension ($z$ goes from $-\infty$ to $+\infty$) while in the case of the axisymmetric lenses the focus forms in a finite volume because in this case one dimension is bounded ($\theta$ goes from 0 to $2\pi$). Then, the symmetry matching in axisymmetric structures is revealed of fundamental relevance to increase the focusing properties of the full source-lens system in practical situations.

In the homogenization limit, it is possible to obtain the acoustic impedance of the lens material as a function of the filling fraction as follows,
\begin{equation}
Z_{eff}(r)=\frac{\sqrt{1+ f(r)}}{1-f(r)}{Z_h}. \label{impedance}
 \end{equation}
where $Z_h=\rho_{h}c_{h}$ is the acoustic impedance of the host medium\cite{Cervera02} (in the current work, air, $\rho_h=1.29$ kg/m$^3$ and $c_h=343$ m/s). In our design, the maximum impedance contrast appears in the center of the lens ($f(r=0)=\pi/4$) and its value is $Z(0)=6.2Z_h$. The impedance profile is governed by the refraction index profile, therefore presenting a decrease of the impedance along the radial coordinate. This impedance profile assures that the acoustic waves are strongly refracted and weakly reflected, which reinforces the high sound focusing obtained by the axisymmetric GRIN lens presented here.

In this work we have designed an axisymmetric GRIN lens presenting a geometry matching with the source. To do that, we have built a system made of rigid toroidal scatterers embedded in air,  by varying the filling fraction in the radial plane in order to produce a hyperbolic secant profile. The ray-tracing of the paraxial approximation, the effective fluid medium approximation, the numerical prediction of FEM and the experimental results, all are in good agreement showing enhanced focusing properties never observed before in this kind of GRIN lenses. Sound amplifications of 8.24 dB have been observed in the focusing spot by our axisymmetric GRIN lenses. This macroscopic lens, due to its axial symmetric design and the geometry matching with most of the acoustic sources could be the motivation for several applications in science and technology ranging from aeroacoustics to microfluidics or ultrasound therapy.

\begin{acknowledgments}
The work was supported by Spanish Ministry of Science and Innovation and European Union FEDER through projects FIS2011-29734-C02-01 and -02 and PAID 2012/253. V.R.G. is grateful for the support of post-doctoral contracts of the UPV CEI-01-11.
\end{acknowledgments}

%\bibliography{Axisymmetric_2}


\end{document}